\documentclass[a4paper,preprintnumbers,twocolumn,floatfix,showpacs,nofootinbib]{revtex4}
\usepackage{psfrag}
\usepackage{subfigure}
\usepackage{color}
\usepackage{mathrsfs}
\usepackage{graphicx}
\def\be{\begin{equation}}
\def\ee{\end{equation}}
\def\bea{\begin{eqnarray}}
\def\eea{\end{eqnarray}}

\begin{document}

\title{The holographic bound in the scalar-tensor and $f(R)$ gravities}

 \author{J. T. Firouzjaee}
\affiliation{ School of Physics and School of Astronomy, Institute for Research in Fundamental Sciences (IPM), Tehran, Iran }
 \email{j.taghizadeh.f@ipm.ir}


\begin{abstract}
The holographic bound has been extended to the different theory of gravities such as scalar-tensor gravity and $f(R)$ gravity according to the Noether charge definition of the entropy for a black hole surface. We have introduced some popular examples of the flat FRW cosmology in order to investigate holographic bound in scalar-tensor and $f(R)$ gravity. Using the holographic bound, we put an additional constraint on the scalar-tensor gravity and $f(R)$ gravity parameters.  We also discuss about the transformation from Jordan frame to Einstein frame.
\end{abstract}


\maketitle
\section{Introduction}

Forty years ago Bekenstein made the primary suggestion that black holes have an entropy equal to a quarter of their surface area in Planck units. He also conjectured a holographic Bound on the number of localized degrees of freedom in a gravitational system \cite{bekenstein}.
The extension of this statement to more general situations leads to the holographic principle  \cite{thooft, susskind95}. Along this way,  Fischler and Susskind \cite{fischler} generalizes and refines a cosmological bound. Subsequently, the holographic bound has been extended with a more covariant formulation \cite{bousso99}. There were many attempts for extending the holography bound in the different theory of gravities such as  Gong \cite{gong} who has extended the holographic bound to the Brans-Dicke cosmology.
There was also another attempt \cite{verlinde} showing that the closed FRW equation  may be
rewritten so as to describe the universe entropy in terms of total energy and Casimir energy. Surprisingly, it turns out that the corresponding formula has correspondence with
the Cardy formula (Cardy-Verlinde formula) for the entropy of a two-dimensional conformal field theory. This formula may
be written as a dynamical entropy bound. By the way, the modified $f(R)$ gravity modeled as an effective fluid and construct the corresponding Cardy-Verlinde formula for it \cite{odinsov}.

In the other side, the dark energy problem has led many attentions to modified theory of gravity. Cosmic acceleration can either be explained by introducing
large amounts of dark energy or considering modifications
to gravity such as the addition of a suitable
function $f(R)$ of the Ricci scalar to the Einstein-Hilbert
action \cite{reviewf(r)}.
Much attentions have been drawn by scalar fields in studies
of the early time universe. A variety of scalar potentials has
been investigated and a number of accelerating inflationary
cosmology have been advocated. On the other hand, alternative approaches for the dark energy problem can be pursued in the scalar-tensor gravity \cite{elizalde04}. A good review for the quintessence model is \cite{scalartensor}.
The prototypical scalar-tensor alternative to general
relativity is Brans-Dicke theory.

To construct the holographic bound in the modified theory of gravities, we should have more  profound look on the entropy of the black holes in these theories. To this aim, Wald and Visser \cite{waldvisser} extend the black hole entropy to the more general theory of gravities which plays a basic role in the holographic bound. Subsequently, the entropy of black
holes in generalized dilaton theories and  in theories with Lagrangians that depend on
an arbitrary function of the Ricci tensor were also examined \cite{alex}. The difference between the form of the entropy in the Einstein gravity and the modified gravity leads us to revisit the black hole area law and the horizon definitions in the modified gravity. By the way, I examined the holographic bound in generalized theories of gravity using the Noether charge definition for the black hole entropy.

The paper is organized as follows. In section II, I extend the holographic bound for general theory of gravity. Section III is devoted to the Holographic Bound in Brans-Dicke Cosmology and  scalar-tensor gravity. We investigate the holographic bound in $f(R)$ cosmology in section IV.
 Section V is devoted to the discussion about the transformation from the Jordan frame to Einstein frame. Section VI contains a discussion and the conclusions.

\section{Holographic bound for general gravity theories}

In this section, the holographic entropy bound in the Susskind process will be  investigated.
Let us look at a scalar-tensor gravity with the Lagrangian density,
\be
{\mathscr{L}} =
F(\phi)\frac{R}{16\pi} +
\mathrm{other\hspace{0.1cm}terms\hspace{0.1cm}independent\hspace{0.1cm}of\hspace{0.1cm}Riemann}
\ee
It has been discussed \cite{alex} that the non-negative entropy in the closed surface, $A$, in scalar-tensor theory for spherically symmetric case on the black horizon is
\be
S =
\frac{F(\phi)A}{4}
\ee

If we consider a well-separated, non-interacting components which
made of some matter components $C_i$ and some black holes with
horizon area $ A_i$, the total entropy of this system is,
\be S_{total}^{initial}=S_{matter}+S_{BH} \ee
which $S_{matter}=\sum S(C_i)$ and $S_{BH}=\sum F(\phi) \frac{A_i}{4}$.
Consider the case that this system interact until a new equilibrium
is established.  The generalized second law (GSL) states, $S_{total}^{final} \geq
S_{total}^{initial}$ , if $S_{total}^{final}$ be the final entropy of the
system.  Now assume that the final state of the
system be a black hole with the entropy $F(\phi) \frac{A}{4}$ (Having this form of the entropy leads that the entropy bound formalism be invariant by the conformal transformation from the Jordan frame to Einstein frame\cite{alex})  this of The GSL
state,
\be S_{matter}+\sum F(\phi) \frac{A_i}{4} \leq F(\phi) \frac{A}{4} \ee

Since $\sum F(\phi) \frac{A_i}{4}$ is obviously non-negative, therefore $S_{matter}
\leq F(\phi) \frac{A}{4}$. This is the new form of entropy bound for
scalar-tensor theory. Similarly for $f(R)$ gravity with the Lagrangian density
\be
{\mathscr{L}} = \frac{f(R)}{16\pi}
\ee
  the entropy bound
will be $S_{matter} \leq f(R)_{,R} \frac{A}{4}$. Note that the nature of the non-negative entropy leads  $ f(R)_{,R} \geq 0$. We should notice
that in scalar-tensor and $f(R)$ gravity, we have the non-constant
gravitational coupling. Therefore, we cannot write the entropy of
different parts separately in general. Generally the the area $A$
can be defined for any two surface but scalar-tensor gravity scalar is a quasi-local
function. Therefore, in this formalism of the entropy bound for the
 scalar-tensor gravity, the scalar field must be calculated at the same place in the boundary with area $A$ .\\
In the case that there isn't any symmetry in the surface, $A$, the holographic bound can be expressed it this way
\be
S_{matter} \leq\int  s_{ab}
\ee
where the $s_{ab}$ is the entropy two form \cite{alex}. In the case of scalar-tensor gravity $ s_{ab} = \frac{F(\phi)}{4} \, \varepsilon_{ab} \,$, and in the case $f(R)$ gravity the entropy two form is $ s_{ab}  =
\frac{f'(R)}{4} \varepsilon_{ab}$ (the $\varepsilon_{ab}$  is the area two form).

It should be stressed that the fact that the requirement of positive black hole entropy simply avoids the appearance of ghost or tachyon fields in the corresponding scalar field theory. Then a negative entropy is simply a footprint of some instabilities in the Einstein frame \cite{negativeentropy}.
As was shown in \cite{elizalde07}, the condition of positive entropy can be used in order to constrain the viability of modified gravity theories.

The covariant holographic bound for scalar-tensor gravity can be stated as follows.
\\
Let $A(B)$ be the area of an arbitrary $D-2$ dimensional spatial
surface $B$ (which need not be closed) and $F(\phi)|_B$ be the
 scalar-tensor gravity scalar value at boundary $B$. $A$ $D-1$ dimensional
hypersurface $L$ is called a light-sheet of $B$ if L is generated by
light rays which begins at $B$, extend orthogonally away from B, and
have non-positive expansion,
\be \theta \leq 0\ee
everywhere on $L$. Let S be the entropy on any light-sheet of $B$.
Then
\be S \leq F(\phi)|_B \frac{A(B)}{4}\label{hb} \ee

The holographic bound for the $f(R)$ gravity can be stated in the same way, but the  $f(R)_{,R} \frac{A}{4}\geq 0$ is an extra assumption which we assume in the classical gravities. On the other side the quantity $G_{eff}= G/f(R)_{,R}$ can be regarded as the effective gravitational coupling strength in analogy to what is done in scalar-tensor gravity \cite{reviewf(r)}. The positivity of $G_{eff}$ or $f(R)_{,R}>0$ is equivalent to the requirement that the graviton is
not a ghost. The positivity assumption of the entropy is closely related to positive  effective gravitational coupling assumption.

\section{ Holographic Bound in Brans-Dicke Cosmology and  scalar-tensor gravity}

The simplest way to incorporate the scalar field as gravitational field is Brans-Dicke theory in which the gravitational coupling constant is
replaced by a scalar field.
The Brans-Dicke Lagrangian in Jordan frame is given by
\begin{equation}
\label{bdlagr}
{\cal L}_{BD}=-\sqrt{-\gamma}\left[\phi{\tilde  R}+
\omega\,\gamma^{\mu\nu}
{\partial_\mu\phi\partial_\nu\phi\over \phi}\right]-{\cal L}_m(\psi,\,
\gamma_{\mu\nu}).
\end{equation}

The homogeneous and isotropic
Friedman-Robertson-Walker (FRW) space-time metric is
\begin{equation}
\label{rwcosm}
ds^2=-dt^2+a^2(t)\left[{dr^2\over 1-k\,r^2}+r^2\,d\Omega\right],
\end{equation}
the above metric can be written as
\begin{equation}
ds^2=-dt^2 + a^2(t)\left[d\chi^2 +f^2 (d \theta^2 + \sin^2
\theta d\phi^2) \right],
\label{1.2b}
\end{equation}
where
\begin{equation}
f=\left\{\begin{array}{lc}
\chi&k=0,\\
\sinh \chi&k=-1,\\
\sin \chi&k=1.
\end{array}\right.
\end{equation}

Based on the FRW metric and the perfect fluid
$T_m^{\mu\nu}=(\rho+ p)\,U^\mu\,U^\nu + p\,g^{\mu\nu}$
as the matter source,
we can get the evolution equations
of the universe from the action (\ref{bdlagr})
\begin{equation}
\label{jbd1}
H^2+{k\over a^2}+H{\dot{\phi}\over \phi}-{\omega\over 6}\left({\dot{\phi}
\over \phi}\right)^2={8\pi\over 3\phi}\rho,
\end{equation}
\begin{equation}
\label{jbd2}
\ddot{\phi}+3H\dot{\phi}=4\pi\beta^2(\rho-3p),
\end{equation}
\begin{equation}
\label{jbd3}
\dot{\rho}+3H(\rho+p)=0.
\end{equation}
Where  $\beta^2=2/(2\omega+3)$. If we are given a state equation for the matter $p=\gamma\rho$, then the
solution to Eq. (\ref{jbd3}) is
\begin{equation}
\label{jbd4}
\rho\,a^{3(\gamma+1)}=C_1.
\end{equation}

For the case $k=0$, we can get the power-law
solutions to the Eqs. (\ref{jbd1}) and (\ref{jbd2})
with the help of Eq. (\ref{jbd4}) \cite{gong},
\begin{equation}
\label{jbd5}
a(t)=a_0\,t^p,\quad \phi(t)=\phi_0\,t^q,
\end{equation}
where
\bea
\label{jbd6}
p={2+2\omega(1-\gamma)\over 4+3\omega(1-\gamma^2)}, \quad
q={2(1-3\gamma)\over 4+3\omega(1-\gamma^2)},\quad \nonumber\\
 -1\le \gamma <  1-{2\over 3+\sqrt{6}/\beta}, \hspace{1.1cm}
\eea
$a_0$ and $\phi_0$ are integration constants,
and $[q(q-1)+3pq]\phi_0=4\pi\beta^2(1-3\gamma)C_1 a_0^{-3(\gamma+1)}$.

To calculate the maximum entropy bound we need to define the apparent horizon surface.
The {\em apparent horizon\/} is defined geometrically as a sphere at
which at least one pair of orthogonal null congruences have zero
expansion.  It satisfies the condition
\begin{equation}
{\dot{a}\over a} = \pm {f'\over a f},
\label{eq-ahcond}
\end{equation}
In the flat case the apparent horizon radius is $ r_{AH}=\frac{1}{\dot{a}}$. 
Let us consider the  light-sheets  which their null generators have negative expansion (for example we consider past ingoing null generator). 
At any time, the spherical area for $ r<r_{H}$ is smaller than the apparent horizon surface area,
$A < A_{AH}$.

The holographic bound can be written,
\begin{equation}
\frac{4S}{ \phi A} \leq \frac{s r_{AH}}{3\phi a(t)^2}=\frac{s t^{1-3p-q}}{3 p \phi_{0}},
\end{equation}
where $s$ is the constant comoving entropy density. This term is maximized by the outermost normal
surface at any given time $t$, on the apparent horizon sphere. The holographic bound in satisfied of $1-3p-q<0$. Therefore, we should have
\be
\frac{-3 \omega \gamma^2+6(w+1)\gamma-3\omega-4}{4+3\omega(1-\gamma^2)} < 0.
\ee

This is the corrected version of the holographic bound which was expressed in \cite{gong}. In the case that we have general relativity limit, $\omega >>1$, we get the standard causal energy condition (Note that, the current solar system observation of Cassini spacecraft requires
that $\omega$ must exceed 40000 \cite{omega}.).

This bound will be held for $\gamma >1+\frac{\sqrt{9+6\omega}+3}{3\omega}$ and $\gamma <1+\frac{\sqrt{9+6\omega}-3}{3\omega} = 1-{2\over 3+\sqrt{6}/\beta}$. Therefore, this holographic bound condition are compatible wtih the condition (\ref{jbd6}).

Let us consider an example of scalar-sensor gravity \cite{elizalde04}. The  scalar-tensor action  can be written as
\begin{eqnarray*}
S&=&{1 \over \kappa^2}\int d^dx \sqrt{-g}\left(F(\Phi)R \mp {1 \over 2}
\partial_\mu \Phi \partial^\mu \Phi - U(\Phi)\right) \nonumber\\
&& + \int d^dx \sqrt{-g}\left( -{1 \over 2}\partial_\mu \chi
\partial^\mu \chi - U(\chi)\right) \ .
\end{eqnarray*}
Here
\[
F(\Phi)={\alpha^2 \over 4|\gamma|}\Phi^2\ ,\quad
U(\Phi)=V_0\left({\alpha \Phi \over 2\sqrt{|\gamma|}}\right)^{4\left(1 - \sqrt{1+{\gamma
\over 3\alpha^2}}\right)}\ .
\]
and  the field $\phi$ is given by
$ \Phi \equiv {2\sqrt{|\gamma|} \over \alpha} e^{{\alpha \over 2}\phi}$.
Here $\alpha$ and $\gamma$ are constant parameters. We define the new variable
\be
\varphi=\phi\sqrt{\alpha^2 + {\gamma \over 3}}.
\ee
\bea
 a=a_{E0}\left({t \over t_0}\right)^{{3 \over 4}\varphi_0^2 - \beta\varphi_0
/ [1 - {\beta \varphi_0 \over 2}]} \nonumber\\
\label{P11c}
\phi={\beta \varphi_0 \over \alpha \left(1 -
{\beta \varphi_0 \over 2}\right)} \ln {t \over t_0}\ .~~~~ \eea
Here
$ \beta\equiv {\alpha \over \sqrt{\alpha^2 +
{\gamma \over 3}}}\ $ and $t_0$ is constant.
The holographic bound is
\bea
\frac{4S}{ F(\Phi) A} \leq \frac{4 s r_{AH}}{3 F(\Phi) a(t)^2} \propto t^{1-{9 \over 4}\varphi_0^2 +\frac{2 \beta\varphi_0 }{ [1 - {\beta \varphi_0 \over 2}]}}.
\label{hbst}
\eea
To satisfying the holographic bound in any time, we get the following condition
\be
1-{9 \over 4}\varphi_0^2 +\frac{2 \beta\varphi_0.
}{ [1 - {\beta \varphi_0 \over 2}]} < 0
\ee
This gives an extra constraint which enable us to limit the on the  $\beta$ and $\varphi_0$ parameters.

\section{ Holographic Bound $f(R)$ Cosmology }

Recently, $f (R)$ theories have been extensively studied in cosmology and gravity such as inflation, dark energy, local gravity constraints, cosmological perturbations, and spherically symmetric solutions in weak and strong gravitational backgrounds. As discussed in section II, It is reasonable to assume that the entropy is non-negative quantity in the classical gravities. \\

 To investigate the holographic bound, we present two famous model in these gravity.
The first example which we will refer to is a model
of the form$ f(R) = R- \mu^{2(n+1)}/R^n$, where $ \mu$ is a suitable
 parameter \cite{carroll}. The Ricci curvature for the flat FRW solution is

\be
R=6\left[ \frac{\ddot{a}}{a}+(\frac{\dot{a}}{a})^2 \right]
\ee
 In this case, the scale factor is assumed to be a generic
power law, $ a(t) \propto t^p$ for the flat case, where
\be
p =\frac{(2n + 1)(n + 1)}{n + 2},
\ee

The  $R$ solution  is, $R=-6p/t^2$. Therefore, $ f'(R)=1+n \mu^{2(n+1)}(-6p)^{-n-1}t^{2n+2}$.
The holographic bound gives
\bea
\label{hbf}
\frac{4S}{ f'(R) A} \leq \frac{4 s r_{AH}}{3f'(R) a(t)^2}=\frac{4 s t^{1-3p}}{3 p f'}\propto t^{-1-2n-3p},
\eea
for $n>-1$.
The holographic bound is satisfied if $-1-2n-3p<0$. This condition will be satisfied for all rang $n>-2$. The best fitting for the astronomical data bounds $n$ in this range \cite{carroll}. Note that there is also another constraint that comes from the positivity of the entropy $ f'(R)=1+n \mu^{2(n+1)}(-6p)^{-n-1}t^{2n+2} > 0$ \cite{elizalde07}.

Another example for the function $f$ is $f(R) \propto R^n$ \cite{Capozziello2003}.  The solution for the flat FRW universe gives the scale factor as $ a(t) \propto t^p$ where
\be
p=\frac{-2n^2+3n-1}{n-2}
\ee
From the Ricci scalar solution we get $f'(R)=(-6p)^{n-1}t^{2-2n}$.

The holographic bound gives
\be
\frac{4S}{ f'(R) A} \leq \frac{4 s r_{AH}}{3f'(R) a(t)^2} \propto t^{-1+2n-3p},
\ee

The holographic bound leads us to this condition $-1+2n-3p=\frac{8n^2-14n+5}{n-2}<0$. Therefore, the holographic bound constraint on $n$ becomes
\be
n < \frac{1}{2}~ ,~~~~  n > \frac{5}{4}
\ee

The observational data limits the $n$  for $-0.450< n < -0.370 $ and $1.366 < n < 1.376 $. Therefore, the holographic entropy bound is satisfied for these intervals of $n$  \cite{Capozziello2003}.\\

The holographic bound condition for the FRW model in Einstein gravity leads us to the causal energy condition \cite{bousso02}. Imposing the energy conditions for the effective stress-energy tensor of f(R) gravity is not very meaningful because the effective stress-energy terms coming from the geometry (when we write the field equations of alternative gravities as effective Einstein equations )  violate all the energy conditions generally \cite{faraoni04}.

\section{the holographic bound in Einstein frame}
Here we recall the conformal transformation which transform the quantities from the Jordan frame to the Einstein frame which is like the Einstein general relativity. It has been argued in the literature that, classically, the two frames are physically equivalent \cite{einstein}. Since the black hole entropy in the Einstein frame have the famus form $S=\frac{A}{4}$, the holographic bound will be the same standard holographic bound which was defined by Bousso \cite{bousso02}. Note that the $\phi$ and $f(R)$ act as a field in the Einstein equation in Einstein frame. Now look at the conformal transformation which transfer from  Jordan frame to Einstein frame;
\be
g_{ab}= \Omega^2 \gamma_{ab}.
\ee
In this equation  $\Omega=\sqrt{ F(\Phi)}$ and $\Omega=\sqrt{ f'(R)}$ in the case of the scalar-tensor gravity and   f(R) theory respectively.
Now consider the FRW cosmology in the Einstein frame;
\begin{equation}
ds^2=-d\tau^2 + R^2(t)\left[d\chi^2 +f^2 (d \theta^2 + \sin^2
\theta d\phi^2) \right],
\label{1.4b}
\end{equation}
The holographic bound must be written,
\begin{equation}
\frac{4S}{ A} \leq \frac{s r_{AH}}{3 R(t)^2}
\end{equation}
where  the apparent horizon radius is $ r_{AH}= 1/\frac{d R(t)}{d\tau}$. Putting  the $d\tau = \sqrt{ F(\Phi)} dt$ and $R(t)= \sqrt{ F(\Phi)} a(t)$  in the above equation, we exactly get the holographic bound for scalar-tensor gravity (\ref{hbst}). Similarly if we insert the $d\tau = \sqrt{ f'(R))} dt$ and $R(t)= \sqrt{ f'(R)} a(t)$ in the above equation, we get the holographic bound for scalar-tensor gravity (\ref{hbf}). Therefore the holographic bounds in these two frame are equivalent and express the same equation.

\section{discussion}

In this paper we have proposed a new version of the holographic bound for scalar-tensor gravity and $f(R)$ theory of gravity. The key point for this derivation is the definition of the entropy for a black hole surface in these theory of gravities \cite{waldvisser}. I have discussed about the holographic aspects of flat FRW cosmology for these gravities. As consequence of holographic bound, some free parameters in these are constrained even in some case we can rule out some models in scalar-tensor and $f(R)$ gravity. 
As we saw, the transformation from the Jordan frame to Einstein frame gives the  standard holographic bound and the holographic bounds in these two frame are equivalent and express the same equation.

\section{acknowledgment}
I would like to thank Alex B. Nielsen for helpful discussion on drafts of this text.

\end{document}